\documentclass[a4paper,twocolumn,prl,showpacs,showkeywords,aps,nofootinbib]{revtex4}
\usepackage{graphicx}
\usepackage{amsmath}
\usepackage{bm}
\newcommand{\beq}{\begin{equation}}
\newcommand{\eeq}{\end{equation}}
\newcommand{\beqa}{\begin{eqnarray}}
\newcommand{\eeqa}{\end{eqnarray}}
\def\ra{\rangle}
\def\la{\langle}

\begin{document}
\title{Multiple Schr\"odinger pictures 
and dynamics 
in shortcuts to adiabaticity}
\author{S. Ib\'a\~{n}ez$^{1}$}
\author{Xi Chen$^{1,2}$}
\author{E. Torrontegui$^{1}$}
\author{A. Ruschhaupt$^{3}$}
\author{J. G. Muga$^{1}$}
\affiliation{$^{1}$Departamento de Qu\'{\i}mica-F\'{\i}sica,
UPV-EHU, Apartado 644, 48080 Bilbao, Spain}

\affiliation{$^{2}$Department of Physics, Shanghai University,
200444 Shanghai, P. R. China}

\affiliation{$^{3}$Institut f\"ur Theoretische Physik, Leibniz
Universit\"{a}t Hannover, Appelstra$\beta$e 2, 30167 Hannover,
Germany}


\begin{abstract}
%
A Schr\"odinger 
equation  
may be transformed by unitary operators into dynamical equations 
in different interaction pictures which share with it a common physical frame,
i.e., the same underlying interactions, processes and dynamics.  
In contrast to this standard scenario, other relations are also possible, such as a common interaction-picture dynamical equation corresponding to several Schr\"odinger equations that represent different physics. This may enable us to design alternative and feasible experimental routes for operations that are a priori difficult or impossible to perform.  
The power of this concept is exemplified by engineering Hamiltonians 
that improve the performance or make realizable several shortcuts to adiabaticity.
%
\end{abstract} 
%
\maketitle
%
%
%
%
{\it Introduction.---} Schr\"odinger, Interaction, or Heisenberg ``representations'' or ``pictures'' of 
a quantum system are linked to each other by unitary transformations that guarantee their formal equivalence.
Changing the picture may be viewed as a change of basis,
so in principle the same information can be extracted from any of them.
This fundamental equivalence is compatible with distinguishing features, 
for example with respect to their usefulness
({\it i}) to calculate, approximate, or modify the dynamics, and ({\it ii}) to describe the system, its properties, and their closeness or otherwise to 
common language or classical notions.  
The Schr\"odinger picture (SP)
is often privileged as the primary description,  
representative of the physical or experimental setting, whereas the multiple interaction pictures (IP) have  the connotation of auxiliary mathematical constructs to facilitate the calculations.
The standard relation among them 
is schematically depicted in Fig. 1(a), where each node 
may represent the dynamical equations (DE) for state vectors, the Hamiltonians, or  the state vectors themselves. The external box means that they all represent the same common underlying physics: the same interactions and external forces, and the same system dynamics (``physical frame'' for short). 

%
%
\begin{figure}[t]
\begin{center}
\includegraphics[width=.6\linewidth]{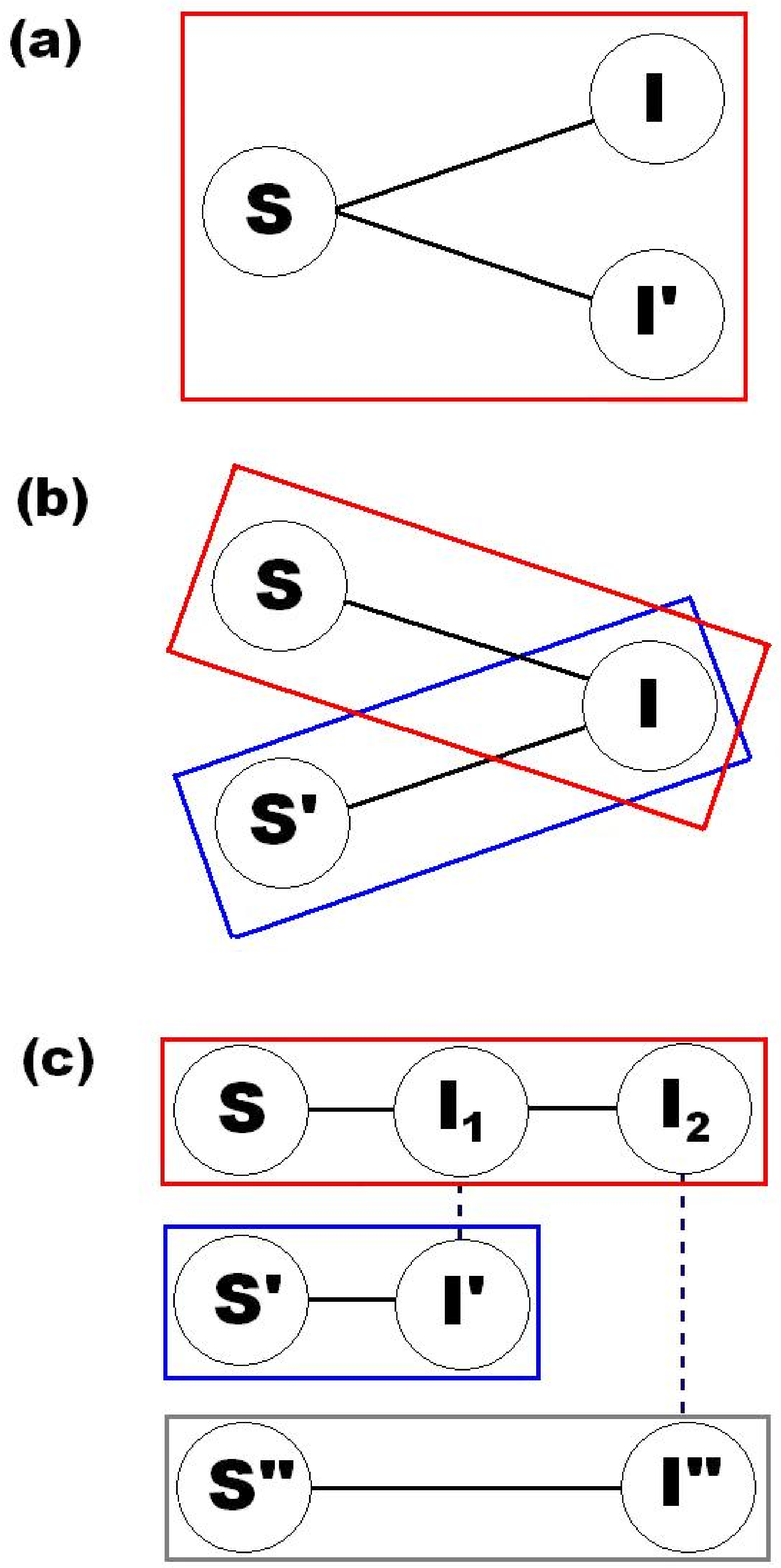}
\end{center}
\caption{\label{fig1} (color online). Schematic relation between different Schr\"odinger and interaction picture dynamical equations. Each node may also 
represent the Hamiltonians or the states. The rectangular boxes (frames) enclose nodes that represent the same underlying physics. The solid lines are unitary relations for the linked states, and 
the dashed lines represent a non-unitary addition of a term to the Hamiltonian. 
}
\end{figure}
%
In sharp contrast to the common wisdom scenario we have just described, 
we propose in this letter alternative relations such as the 
ones in Figs. 1(b), 1(c) or more complex combinations, 
where the nodes may belong to different 
physical frames, as a way to design alternative and feasible experimental routes for operations that are a priori difficult or impossible to perform. 
This will be applied in particular to engineer Hamiltonians 
which improve or make feasible ``shortcuts to adiabaticity'' \cite{Ch10}. 
Adiabatic processes are very common
and useful in laboratories, but their intrinsic slowness
imposes limitations. 
The design of alternative fast routes is
an active research field of interest in cold atom physics,
nuclear magnetic resonance, quantum information processing, 
and beyond the quantum domain, e.g. to couple
different devices in optical communications \cite{modetransf}. 
Shortcuts designed on paper, may or may not be feasible in practice,
so the possibility to generate multiple physical frames proves useful.  
Our first benchmark problem is the acceleration of adiabatic population inversion in a two-level system, in itself a phenomenon of broad interest from NMR applications \cite{NMR} to quantum information  \cite{Vitanov}. We shall point out several techniques to eliminate undesired Hamiltonian terms, possible improvements to reinterpreted experiments, and also feasible alternatives to inversion schemes that required cumbersome level-shift engineering 
or multiple fields \cite{Chen10b}.   
We shall finally show how to overcome the difficulties to implement ``counterdiabatic'' terms and perform fast expansions or compressions of cold atoms without final excitation \cite{harmo}.

{\it{Multiple Schr\"odinger pictures.---}}
%
To explain the multi-frame schemes of Figs. 1(b) and 1(c) we 
need first to  
review some basic equations and notation.            
SP and IP states are related by a unitary transformation, 
%
$|\psi_I\ra=U^\dagger|\psi_S\ra,\;\;\;|\psi_S\ra=U|\psi_I\ra$,
%
and evolve according to 
\beq
i\hbar\partial_t|\psi_S\ra=H|\psi_S\ra,\;\;i\hbar\partial_t|\psi_I\ra=H_I|\psi_I\ra,
\label{IPEE}
\eeq
where $H$ is the Hamiltonian in the SP and 
\beq
H_I=U^\dagger(H-K)U,\;\;{\rm with}\;\; K=i\hbar \dot{U}U^\dagger,
\label{kop}
\eeq 
the corresponding IP Hamiltonian.  
Note that $H$, $H_I$ and $K$ may be generally time-dependent. 
Similar relations hold for a unitary operator $U'$ which defines an interaction picture $I'$.

%
As in Fig. 1(b), an interaction picture DE   
may be related unitarily to two or more Schr\"odinger equations with Hamiltonians that represent different experimental settings and external interactions.   
There is no contradiction 
with the equivalence discussed above  when we pay attention, not only to the DEs but to the observables as well. A given picture is fully characterized by both the DE and the operators for the observables. Thus Fig. 1(b) admits several interpretations depending on the treatment given to the observables:   
If the observables are transformed, from $B_S$ in S to $B_{S'}={U'} U^\dagger B_S U U'^\dagger$ in $S'$, we will get the same expectation values from two different systems and dynamics performing in general rather different measurement operations.  
If instead one runs the same measurements in $S$ and $S'$    
on the same, untransformed observables $B_S$, the expectation values will differ in general,   
and their IP representatives would be either $U^\dagger B_S U$ 
or ${U'}^\dagger B_S U'$, sharing in any case a common IP-state dynamics.   
It may also occur that, having applied the transformations, the equality $B_{S'}=B_{S}$ holds for specific (but not for all) observables, as in several examples below.  
 


Another multi-frame scheme is depicted in Fig. 1(c). In the upper box
a Schr\"odinger node 
is related unitarily to a
first IP-node, $I_1$, linked also unitarily to a second one, $I_2$.
The two consecutive IP Hamiltonians may be modified or perturbed, e.g. by the addition of some terms (dashed lines).  
This changes the physics into $I'$ and $I''$, each in a different physical frame 
(middle and lower horizontal boxes)
with corresponding Schr\"odinger dynamics.      
In the example below the consecutive IPs are generated by means of adiabatic and superadiabatic 
iterations \cite{Garrido, Berrysa}, and the addition of a ``counterdiabatic'' term in the Hamiltonian is performed so as to avoid transitions, canceling out the $K$
in Eq.(\ref{kop}) \cite{Rice08}. This enables us to accelerate slow processes without inducing any final excitation.      


     
%
%
%

{\it Superadiabatic iterations and counterdiabatic corrections.---}
Our starting model Hamiltonian is 
%
%
%
\beq
H_j(t)=\left(
\begin{array}{cc}
Z_j(t) & X_j(t)-iY_j(t)
\\
X_j(t)+iY_j(t)& -Z_j(t)
\end{array}
\right)
\label{basicH}
\eeq
for $j=0$ and $Y_0=0$, i.e., $H_0=X_0\sigma_x+Z_0\sigma_z$ in terms of Pauli 
matrices. 
It could represent several physical systems such as a spin in a magnetic field,
a two-level atom, or a condensate in the bands of an accelerated optical lattice \cite{Oliver0,Oliver}. 
In the later case, $X_0$ may be controlled by the trap depth, $Z_0$ by the lattice acceleration \cite{Oliver0}, and a $Y_0$ component could in principle be implemented by a second shifted lattice \cite{Oliver}. The index $j$ will be used later to define a series of IP Hamiltonians in successive iterations. The Hamiltonian evolution or ``trajectory'' is here specified by the Cartesian coordinates $X_j$, $Y_j$, $Z_j$.   Later we shall also use the corresponding polar, azimuthal, and radial spherical coordinates, $\Theta_j$, $\Phi_j$, 
and $R_j$. 
The radius may or may not be constant with respect to time, so the trajectory is not generally on a sphere.         
The Hamiltonian matrices are  expressed in the ``bare basis'' of the two-level system, 
$|1\rangle = (\tiny{\begin{array} {c} 1\\ 0 \end{array}})$, $|2\rangle = (\tiny{\begin{array} {c} 0\\ 1 \end{array}})$. 
We assume that this is also the eigenbasis of $H_j(t)$, or very close to it
for computational purposes, at initial and final times $t=0$ and $t=t_f$.    

Focusing by now on $j=0$, an adiabatic population inversion is achieved with $H_0$ by varying slowly
$X_0$ and $Z_0$ so that the resonance is crossed at $Z_0=0$, and the eigenvectors of $H_0$ interchange their character. Different schemes, such as 
Landau-Zener (LZ), Allen-Eberly \cite{AE}, and others, may be followed to specify the time-dependences.  
The first IP that we shall consider depends on the adiabatic basis $\{|n_0(t)\ra\}$
that diagonalizes $H_0(t)$ 
and keeps $K(t)$ non-diagonal.
Specifically we use $U=A_0$,   
\beq
A_0(t)=\sum_{n=1,2} |n_0(t)\ra\la n_0(0)|, 
\label{a0}
\eeq
where we assume that the $|n_0(0)\ra$ coincide with the bare basis. 
The corresponding $K$ operator in Eq. (\ref{kop}) is denoted as $K_0$. Also, 
$|\psi_{I_1}\ra=A_0^\dagger|\psi_S\ra$ and $H_1=A_0^\dagger(H_0-K_0)A_0$.   
Constructing $\{|n_0(t)\ra\}$ requires a proper choice of phases. From an arbitrary adiabatic
basis $\{|n_a(t)\ra\}$ that diagonalizes $H_0$ with eigenvalues $E_n(t)$, 
\beq
|n_0(t)\ra=e^{i\gamma_n}|n_a(t)\ra,
\eeq
where $\gamma_n=i\int_0^t dt' \la n_a(t')|\dot{n}_a(t')\ra$ is the geometric phase
and the dot denotes time derivative. 
This phase choice is privileged since $\la n_0(t)|\dot{n}_0(t)\ra=0$. It 
also  makes $K_0$ non-diagonal and 
minimizes its norm \cite{Mes,Rice08}. 
In terms of the polar angle $\Theta_0$,   
\beq
K_0=i\hbar\dot{A}_0A_0^\dagger=\hbar(\dot{\Theta}_0/2)\sigma_y.
\eeq
%
The adiabatic approximation  neglects $K_0$ in the IP Hamiltonian $H_1$ 
to trivially solve  
\beq
\label{IPA}
i\hbar\partial_t |\psi_{I_1}\ra=A_0^\dagger H_0 A_0 |\psi_{I_1}\ra, 
\eeq
an uncoupled system in the bare  
basis. 
Alternatively one may add $A_0^\dagger K_0A_0$ to $H_1$  
\cite{Rice03,Rice05,Rice08,Berry09,Chen10b}.
The effect is to cancel any coupling so that Eq. (\ref{IPA}) 
becomes exact rather than an approximation. In the corresponding SP $S'$, see the middle box in Fig. 1(c), 
this amounts  to adding the counterdiabatic term $H_{cd}^{(0)}:=K_0$ to $H_0$.  
$H_0+K_0$
preserves the populations of the 
approximate adiabatic dynamics even for short process times. 

In a new iteration, and similarly for higher orders, 
we write $H_1=A_0^\dagger(H_0-K_0)A_0$ in the form of Eq. (\ref{basicH}), $j=1$,  and 
diagonalize it to produce a ``superadiabatic'' basis 
$\{|n_1(t)\ra\}$, and the transformation 
\beq
A_1=\sum_{n=1,2} |n_1(t)\ra\la n_1(0)|.
\label{a1}
\eeq    
As before we assume that this basis coincides at the boundary times with the bare basis and that
$K_1=i\hbar\dot{A}_1A_1^\dagger$ is 
non diagonal in $\{|n_1(t)\ra\}$.     
$A_1$ produces a new IP ($I_2$ in Fig. 1(c)) with   $|\psi_{I_2}\ra=A_1^\dagger|\psi_{I_1}\ra$, and Hamiltonian $H_2=A_1^\dagger(H_1-K_1)A_1$.    
$K_1$ 
can be either neglected 
to produce a superadiabatic approximation, or   
canceled by adding  a counterdiabatic term. In the corresponding SP
($S''$ in Fig. 1(c)) 
the Hamiltonian becomes $H_0+H_{cd}^{(1)}$, where $H_{cd}^{(1)}=A_0 K_1 A_0^\dagger$ \cite{Rice08}.  
In that manner a different shortcut Hamiltonian is created. 
For our reference Hamiltonian $H_0$ with $Y_0=0$, and using polar angles 
for $H_0$ and $H_1$, 
\beq  
H_{cd}^{(1)}=\hbar(\dot{\Theta}_1/2)(\cos\Theta_0\sigma_x-\sin\Theta_0\sigma_z),
\eeq
if $\dot{\Theta}_0<0$. 
The absence of a $Y\sigma_y$ component, like in $H_0$ and unlike $H_{cd}^{(0)}$, 
is in some applications a practical advantage. For example, in an optical lattice implementation of the two-level system 
only one optical lattice is required 
\cite{Oliver}; and in a two-level atom realization, discussed below, it 
avoids the application of a second laser \cite{Chen10b}.  
One more advantage of the superadiabatic shortcut is that $H_{cd}^{(1)}$ is less intense (it has a smaller norm)
than $H_{cd}^{(0)}$ \cite{Rice08}.  
Alternative eliminations of $\sigma_y$ are discussed next. 

{\it $Z$-axis rotation.---}
Starting from the SP dynamical equation with $H_0+H_{cd}^{(0)}$ 
that we write now in the form 
\beq
H_0+H_{cd}^{(0)}=
\left(
\begin{array}{cc}
Z_0&Pe^{-i\phi}
\\
P e^{i\phi}& -Z_0
\end{array}
\right),
\label{oldH}
\eeq
where $\phi=\arctan (\hbar\dot{\Theta}_0/2X_0)$, 
$0\le\phi<2\pi$,  
and $P=[X_0^2+(\hbar\dot{\Theta}_0/2)^2]^{1/2}$,
we may apply the
transformation \cite{Berry90}
\beq
U_z=\left(
\begin{array}{cc}
e^{-i\phi/2}&0
\\
0& e^{i\phi/2}
\end{array}
\right),
\eeq 
which amounts to a rotation about the $Z$ axis by    
$\phi$. Notice that because $U_z$ is diagonal in the bare basis,
the bare-state populations do not change from the SP to the IP. 
In the corresponding IP, and with $K_z=i\hbar\dot{U}_zU_z^\dagger$, the interaction Hamiltonian becomes  
\beq
U_z^\dagger(H_0+H_{cd}^{(0)}-K_z)U_z=
\left(\!\!
\begin{array}{cc}
Z_0-\hbar \dot{\phi}/2&P
\\
P & -Z_0+\hbar \dot{\phi}/2
\end{array}
\!\!\right),  
\label{newH}
\eeq
%
without $Y\sigma_y$ component. It  can be 
realized directly in the laboratory 
and we may treat it as well as a SP Hamiltonian
linked to the $I'$ Hamiltonian  
$A_0^\dagger H_0 A_0$,    
a common IP node for the two SP Hamiltonians in Eqs. (\ref{oldH})
and (\ref{newH}), 
connected via $A_0$ and $U_z^\dagger A_0$ respectively. 
This Hamiltonian trio and the corresponding dynamical equations constitute a neat 
example of the dual physical frame scheme of Fig. 1(b).         
Eq. (\ref{newH}) provides an alternative  shortcut path, that guarantees the same bare-state populations than $H_0+H_{cd}^{(0)}$, and indeed it has  
been implemented experimentally for a condensate on an accelerating 
lattice \cite{Oliver}, to avoid the realization of a 
$\sigma_y$ term with a second optical lattice.  
The transition from Eq. (\ref{oldH}) to (\ref{newH}) was justified based on properties specific to the optical lattice setting in \cite{Oliver}.
In fact the elimination of $\sigma_y$ in the Hamiltonian 
can be done formally for any physical realization, and its usefulness will depend on the feasability to implement the modified $X$ and $Z$ terms, 
demonstrated for a condensate on an accelerating lattice, but more involved  
for a two-level atom in an oscillating field, see  below.     

\begin{figure}[t]
\begin{center}
\hspace*{-0.4cm}\includegraphics[width=.74\linewidth]{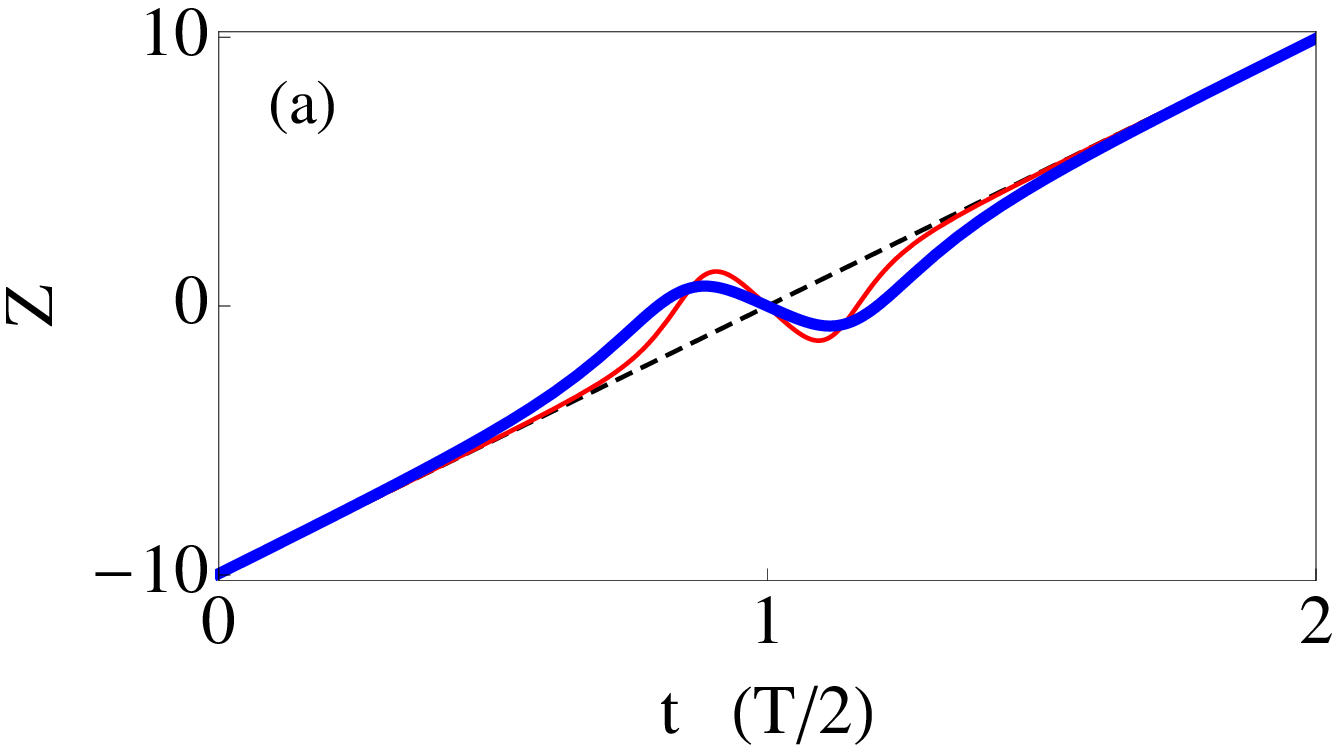}
\includegraphics[width=.7\linewidth]{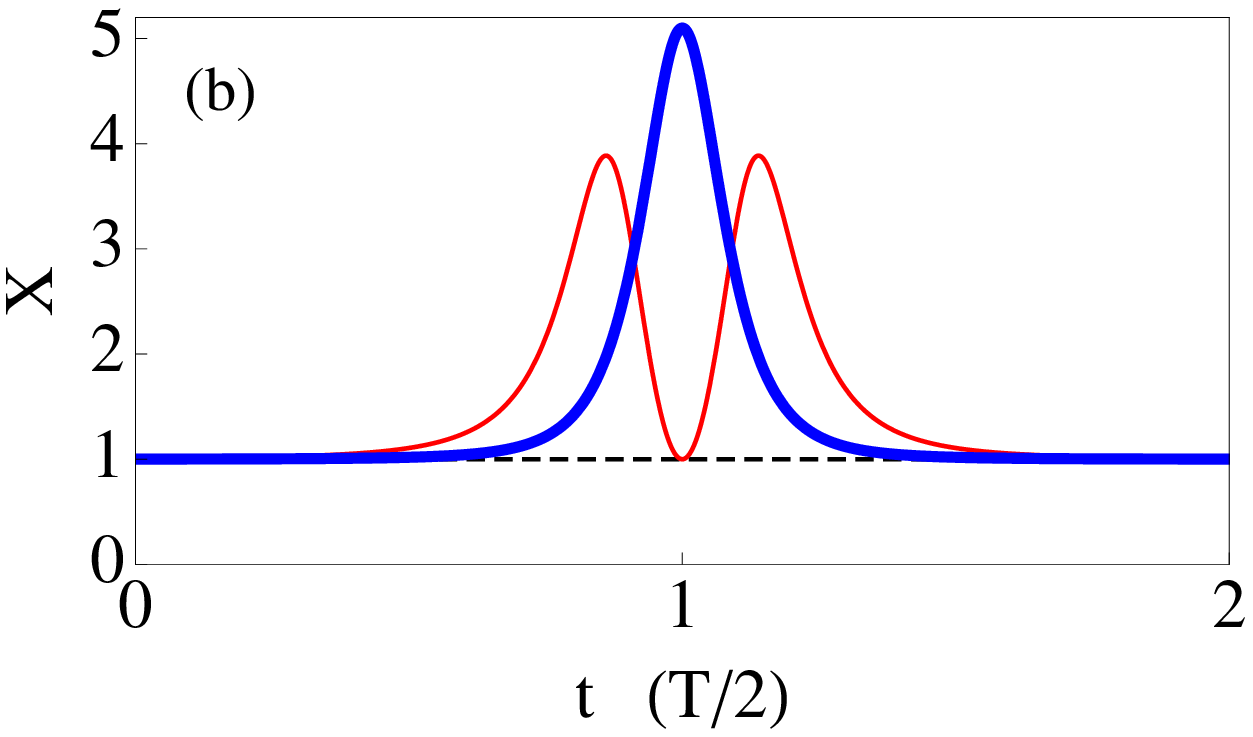}
\end{center}
\caption{\label{fig2} (color online). $Z$ and $X$ Hamiltonian components for 
$H_0$ (dashed lines); 
$z$-axis rotation Hamiltonian (\ref{newH}), $U_z^\dagger(H_0+H_{cd}^{(0)})U_z$ (thick blue solid lines); superadiabatic+counterdiabatic method Hamiltonian 
$H_0+H_{cd}^{(1)}$ (thin red solid line). 
$Z_0(t) =  \alpha  (t-T/2)$, and 
$\alpha=-10$, $T=20/|\alpha|$, in units $\hbar=1$, $X_0=1$.    
\label{GaOm}}
\end{figure}

\begin{figure}[t]
\begin{center}
\includegraphics[width=.7\linewidth]{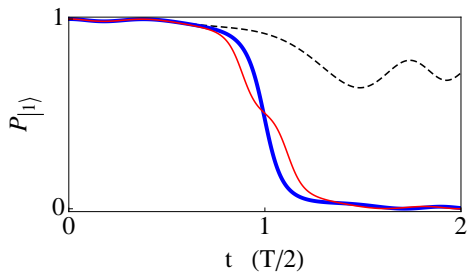}
\end{center}
\caption{\label{fig3} (color online). Populations of the bare state $|1\ra$ 
corresponding to the Hamiltonians depicted in Fig. \ref{GaOm}.
Same line codes as in that figure. Time in units of $T/2$. \label{Popu}}
\end{figure}

The approach based on a $Z$-rotation is compared in Figs. 2 and 3 
with the one based on adding to $H_0$ the  counterdiabatic term $H_{cd}^{(1)}$
with the LZ scheme for $H_0$ (i.e. a constant $X_0$ and a linear in time 
$Z_0$). Fig. \ref{fig2} shows the Hamiltonian matrix elements and Fig. \ref{fig3} the populations of $|1\ra$.
$H_0$ and the corresponding population are also shown as a reference. 
The process time is chosen to be short so that adiabaticity and population inversion fail for this Hamiltonian.   
Instead, the two shortcuts 
lead to perfect population inversion.   
Their $Z$ components are similar,
but the $X$ components have a rather different structure. A possible advantage of 
the superadiabatic+counterdiabatic approach 
using $H_0+H_{cd}^{(1)}$ is the 
smaller value of the $X$-maximum, which reduces amplitude noise and the field intensity.   
%
%
%

%
%
%
%
%
{\it{Two-level atoms}.---}
In quantum optics, Eq. (\ref{basicH}), with $X_0=\hbar\Omega_R/2$, $Y_0=0$, 
and $Z_0=-\hbar\Delta/2$, represents a rotating frame IP
Hamiltonian for a two-level atom in an oscillating field 
with angular frequency $\omega(t)=\omega_0-\Delta(t)$, where $\omega_0$ is the (angular) transition frequency, 
$\Omega_R$ the (on-resonance) Rabi frequency and $\Delta$ the detuning,  
after having applied the electric dipole and rotating wave approximations (RWA). 

For $K=K_L=-[\hbar\omega(t)/2]\sigma_z$ and 
$U=U_L=\exp[-(i/\hbar)\int_0^t K_L(t') dt']$ (As $U_L$ is diagonal in the bare basis, the populations
are the same in interaction and Schr\"odinger pictures.) the corresponding $S$ (RWA) Hamiltonian is 
\beq
K_L+U_L H_0 U_L^\dagger=
\frac{\hbar}{2}\left(
\begin{array}{cc}
-\omega_0&\Omega_R e^{i\theta} 
\\
\Omega_Re^{-i\theta}& \omega_0
\end{array}
\right),
\label{firstH}
\eeq
where $\theta(t)=\int_0^t \omega(t') dt'$.    
We can read from it the time dependent intensity, proportional to $\Omega_R^2$, the  frequency of the field that has
to be applied, $\dot\theta/(2\pi)$, and the atomic transition frequency $\omega_0/(2\pi)$.  
If we start instead with $H_0+H_{cd}^{(0)}$ in the IP DE and apply the same transformation as before, we get the SP Hamiltonian 
\beqa
K_L&+&U_L (H_0+H_{cd}^{(0)}) U_L^\dagger
\nonumber\\
&=&
\frac{\hbar}{2}\left(
\begin{array}{cc}
-\omega_0&(\Omega_R-i\dot{\Theta}_0) e^{i\theta} 
\\
(\Omega_R+i\dot{\Theta}_0)e^{-i\theta}& \omega_0
\end{array}
\right),
\eeqa
which requires in principle two fields dephased by $\pi/2$ sharing a common time-dependent frequency but with different time-dependent intensities \cite{Chen10b}. 
A $Z$-rotation may also be applied  but realizing the result is now complicated 
due to the time-dependence of the diagonal components. i.e. of the transition frequency.
In this physical context that dependence would 
imply  time dependent level-shift engineering with an additional laser. It is thus advisable to find 
alternative, simpler realizations of the shortcuts.
      
$H_0+H_{cd}^{(0)}$ is not the only Hamiltonian that drives the populations along the ones of the adiabatic approximation for $H_0$.
There is a whole family of them using different phases for the adiabatic base states, the simplest one being $H_{cd}^{(0)}$ itself   
\cite{Berry09,CTM}. Note that $A_0^\dagger (H_{cd}^{(0)}-K_0)A_0=0$ so the state does not  move at all in the corresponding IP, whereas in the DE driven by $H_{cd}^{(0)}$ the populations will follow the ones for the adiabatic dynamics of $H_0$.  
By contrast, $H_{cd}^{(1)}$ alone is not enough to take the system 
along the superadiabatic path defined by $H_1$. We could still get rid of $H_0$ 
and use as a shortcut to superadiabaticity $H_{cd}^{(01)}:=H_{cd}^{(0)}+H_{cd}^{(1)}$. 
To see why, use $U_{01}=A_0A_1$, and notice that
$U_{01}^\dagger(H_{cd}^{(01)}-i\hbar\dot{U}_{01}U^\dagger_{01})U_{01}=0$.   
In any case $H_{cd}^{(01)}$ is not so interesting for the present application 
as it combines the three Cartesian components.     

Let us now take $H_{cd}^{(0)}=K_0$ as the reference IP 
Hamiltonian and try to implement it with different physical fields 
as in Fig. 1(b).   
Applying $U_L$   
we get the SP Hamiltonian
%
\beq
H_S=K_L+U_LK_0U_L^\dagger=
\frac{\hbar}{2}\left(
\begin{array}{cc}
-\omega(t)&-i \dot{\Theta}_0 e^{i\theta}
\\
i\dot{\Theta}_0 e^{-i\theta}& \omega(t)
\end{array}
\right),
\eeq
which is indeed problematic to realize because the atomic transition 
frequency should be time dependent.  
In other words, a simple IP Hamiltonian  does not necessarily imply a simple experiment.      
To remedy this, keeping  the same simple IP DE, we may use  
instead $U'=e^{-(i/\hbar)\int_0^t K'(t') dt'}$, with $K'=-(\hbar/2) \omega_0\sigma_z$.    
This choice implies now a simple resonant interaction with constant frequency
$\omega_0$ and $S'$ Hamiltonian 
\beq
H_{S'}=K'+U'K_0{U'}^\dagger=\frac{\hbar}{2}
\left(
\begin{array}{cc}
-\omega_0&-i\dot{\Theta}_0 e^{i\omega_0 t}
\\
i\dot{\Theta}_0 e^{-i\omega_0 t}& \omega_0
\end{array}
\right).
\eeq
Other single laser implementations may also be developed by starting instead 
with $H_0+H_{cd}^{(1)}$. The term $H_{cd}^{(1)}$ modifies the detuning and Rabi frequency so that the transformation $U_L$ would lead to an SP Hamiltonian 
with the same structure as Eq. (\ref{firstH}), but with modified laser and
Rabi frequencies. A further alternative to the superadiabatic iterations is the 
``invariants-based inverse engineering approach'' \cite{CTM}. 
 
{\it Trap expansions.---} The final example is a fast harmonic trap expansion, 
or compression, which is receiving much attention because of fundamental and practical implications \cite{Salamon09,Ch10,Nice10,MN10,energy,OCT,Nice11,Adol,EPL11}. The 
reference Hamiltonian is 
$H_h=p^2/(2m)+m\tilde{\omega}^2q^2/2$, where $\tilde{\omega}=\tilde{\omega}(t)$ is the time dependent angular frequency, $m$ the particle mass, and $q$ and $p$ are 
position and momentum operators.  
The corresponding counterdiabatic term to avoid excitations is
\cite{harmo} $H_{cd}^{(0)}=-(pq+qp)\dot{\tilde{\omega}}/(4\tilde{\omega})$, whose direct 
laboratory implementation 
is problematic and was left as an open question \cite{harmo}. This difficulty is overcome by the transformation $U_q=\exp{(i\frac{m\dot{\tilde{\omega}}}{4\hbar\tilde{\omega}}q^2)}$, which eliminates the cross terms; it produces 
from $H_S=H_h+H_{cd}^{(0)}$ the IP Hamiltonian
$H_I=U_q^\dagger(H_S-i\hbar\dot{U}_qU_q^\dagger)U_q=p^2/(2m)+m\tilde{\omega}'^2q^2/2$, where
\beq  \tilde{\omega}'=\left[\tilde{\omega}^2-\frac{3\dot{\tilde{\omega}}^2}{4\tilde{\omega}^2}+\frac{\ddot{\tilde{\omega}}}{2\tilde{\omega}}\right]^{1/2}.
\eeq
This Hamiltonian can actually be realized directly \cite{Nice10,Nice11} 
and considered in a different 
physical frame as an ordinary harmonic oscillator with modified frequency.
To satisfy the scheme of Fig. 1(b) we may apply $U=1$ and regard it as a Schr\"odinger Hamiltonian $S'$, namely $H_I=H_{S'}$. It indeed provides a shortcut with the following properties: starting with a common state at time $t=0$, the spatial densities driven by $H_h+H_{cd}^{(0)}$ and $H_{S'}$ are identical. In fact, by imposing $\dot{\tilde{\omega}}(t_f)=\ddot{\tilde{\omega}}(t_f)=0$ the final state is also equal for both dynamics, even in phase, and the final vibrational state populations coincide with those of a slow adiabatic process.

{\it Discussion.---} We have first proposed schemes for which different interaction and Schr\"odinger picture dynamical equations represent different physical processes
and interactions.
These schemes have been later combined and exemplified to produce better, realizable shortcuts to adiabaticity for population inversion protocols, and for 
expansions and compressions. 
Similar manipulations may be applied as well to 
facilitate or improve shortcuts to adiabaticity for other operations such as  controlled atomic transport \cite{David,Calarco,transport}. In fact the idea of designing the pictures to generate alternative, easier to handle physics, 
is applicable to a plethora of quantum systems, in particular,
in the realms of quantum simulations, quantum control, or quantum information, 
where developing techniques  to externally drive the systems for specific goals is a central objective.

We are grateful to M. V. Berry, M. Demirplak, D. Gu\'ery-Odelin, and O. Morsch for discussions. 
We acknowledge funding by Projects No. GIU07/40, No. FIS2009-12773-C02-01,
No. 60806041, No. 61176118,  
and Juan de la Cierva Program.  
S. I. and E. T. acknowledge financial support from the Basque Government (Grants No.  BFI09.39 and BFI08.151). 

\end{document}